\documentclass[aps,prl,twocolumn,amsmath,amssymb,10pt]{revtex4-2}
\usepackage{amsmath, amsfonts, amssymb, mathrsfs, dsfont}
\usepackage{bm, bbm, booktabs, mathtools}
\usepackage{pifont, amsthm, amscd}
\usepackage{graphicx}

\def\be{\begin{equation}}
\def\ee{\end{equation}}
\def\ba{\begin{array}}
\def\ea{\end{array}}

\usepackage[nopatch=footnote]{microtype}

\begin{document}

\title{Dynamical evolution of quantum mutual information in Schwarzschild spacetime}

\author{Guosen Ma$^{1, \ast}$, Fang Rao$^{1, \ast}$, Jingjing Hua$^{1, \ast}$, Xiaofen Huang$^{1, \dag}$}
\affiliation{
  $^{1}$ School of Mathematics and Statistics, Hainan Normal University, Haikou, 571158, China\\
  $^{\ast}$ These authors contributed equally to this work.\\
  $^{\dag}$ Correspondence to: huangxf1206@163.com
}

\begin{abstract}
Quantum mutual information is a fundamental quantity for characterizing correlations among quantum systems. In this work, we investigate the dynamic evolution of the quantum mutual information based on R\'enyi-2 entropy for three-mode Gaussian states in the background of a Schwarzschild black hole. 
We find that the physically inaccessible mutual information increases with Hawking temperature and eventually saturates, while the physically accessible mutual information exhibits nonmonotonic temperature dependence: it first rises, reaches a peak, then declines toward a finite asymptotic value. 
This nonmonotonic behavior differs from the monotonic degradation typically observed for Gaussian entanglement and steering in curved spacetime. Furthermore, we establish several constraint 
relations governing the distribution of mutual information among subsystems. These findings contribute to understanding the quantumness of quantum correlations for continuous variable in Schwarzschild spacetime.
\end{abstract}

\maketitle

\section{Introduction}

Quantum mutual information quantifies the total correlations, classical and quantum alike, shared between partitions of a composite quantum system, and thus serves as a fundamental tool for tracking how information flows and redistributes in interacting many-body systems~\cite{GroismanPopescuWinter2005,CerfAdami1997,Kumar2017,NielsenChuang2010,Wilde2013,Chen2026QCD}. In relativistic settings, event horizons causally separate field modes into accessible and inaccessible regions due to Bogoliubov transformation. This causal separation makes the quantum correlation of a quantum system perspective particularly well suited to studying information redistribution in curved spacetimes, one of central topic in relativistic quantum information (RQI)~\cite{PeresTerno2004,MannRalph2012}. The Schwarzschild black hole provides the simplest and most extensively studied gravitational background for examining how the Hawking--Unruh effect reshapes mode accessibility and redistributes quantum correlations across the event horizon~\cite{FuentesSchullerMann2005,PanJing2008}.

In discrete-variable quantum systems, relativistic effects on quantum correlations have been extensively studied~\cite{AlsingMilburn2003,Bruschi2010, AlsingFuentesMannTessier2006,HuYu2011}. In Ref.~\cite{PanJing2008}, the authors demonstrated that entanglement is not simply quenched by the black hole, but rather redistributed between physically accessible and inaccessible regions across the horizon. This redistribution paradigm has since been validated in increasingly general settings, including genuine tripartite and tetrapartite entanglement for tripartite quantum states~\cite{WangPanJing2010,XuSongShiYe2014PRD,DaiShenShi2016PRD,Dong2019GHZ,Dong2019Wclass,TorresArenas2019,TorresArenas2019CPB}, and extends to other quantum resources such as nonlocality and coherence~\cite{XuSongShiYe2014PLB,WangTianJingFan2016,Dong2019Pseudo}. Further investigations have examined genuinely multipartite entanglement, as well as quantum steering, in Schwarzschild backgrounds~\cite{AliAlKuwariGhominejad2024, Wu2024,Mi2026}. 


 Gaussian quantum states provide a complementary platform of relativistic quantum information research for continuous variable quantum system, benefiting from an analytically tractable covariance matrix formalism that maps directly onto field modes~\cite{AdessoFuentesEricsson2007,Weedbrook2012}. The continuous variable RQI framework was pioneered by Adesso et al., who analyzed the entanglement sharing structure of two-mode Gaussian states in noninertial frames and showed that acceleration degrades but does not completely eliminate observable entanglement~\cite{AdessoFuentesEricsson2007}. From a channel perspective, Schwarzschild black holes can equivalently be described as bosonic Gaussian channels, establishing a formal link between gravitational physics and Gaussian quantum information theory~\cite{BradlerHaydenPanangaden2015}. 
 
 Methodologically, a unified characterization of quantum, classical, and total correlations in Gaussian states based on R\'enyi-2 entropy was subsequently developed, supplying a systematic toolbox for studying continuous variable correlations in relativistic settings~\cite{AdessoRagyLee2012CQG}. Wang et al. further extended the analysis to curved spacetime, examining Gaussian quantum steering in Schwarzschild backgrounds and revealing an asymmetry between accessible and inaccessible steering that reflects the underlying entanglement redistribution pattern~\cite{WangCaoJingFan2016}. More recently, Wu et al. generalized the study to tripartite systems, investigating tripartite Gaussian steering in Schwarzschild spacetime and identifying sudden death phenomena for specific partitions~\cite{Wu2025}. Complementary work has examined quantum coherence of bosonic fields in noninertial frames and black hole atmospheres, offering additional perspectives on horizon-induced correlation reshaping~\cite{WuZengCao2021CQG,LiuWenWang2026PLB}. Taken together, these studies have built a solid understanding of Gaussian quantum correlations in relativistic settings, yet a unified total-correlation perspective remains underexplored.

Although the above body of work has thoroughly characterized Gaussian entanglement, steering, and coherence in relativistic settings, these resource-specific measures each capture a distinct aspect of horizon-induced correlation reshaping: entanglement and steering probe purely quantum correlations of different strength, while coherence characterizes single-mode quantum superposition. A complete picture of how correlations flow across the event horizon requires a quantifier of total correlations that accounts for both classical and quantum contributions simultaneously.

For discrete-variable GHZ states, Ali et al. have recently taken steps in this direction by examining mutual information near Schwarzschild horizons~\cite{AliAlKuwariGhominejad2024}. For continuous-variable Gaussian systems, however, a systematic analysis of mutual information dynamics, particularly in multipartite settings where nontrivial sharing structures emerge, is still absent from the literature. The R\'enyi entropy of order 2 provides a natural and well-established framework for filling this gap. It is analytically tractable, directly computable from the covariance matrix without full state diagonalization, and constitutes a bona fide measure of total correlations with firm information-theoretic foundations, relating to quantities such as the accessible information and the quantum fidelity. For Gaussian states specifically, it is the standard tool for characterizing quantum, classical, and total correlations within a unified formalism~\cite{Adesso2012,AdessoRagyLee2012CQG,Lenggenhager2025RMI}.

In this paper, we investigate the evolution of R\'enyi-2 tripartite mutual information for symmetric three-mode squeezed Gaussian states in Schwarzschild spacetime. Going beyond bipartite settings, the tripartite configuration allows us to explore how correlations are distributed among multiple observers and across the event horizon simultaneously, revealing sharing structures that are absent in two-mode systems. We consider a setup where Alice remains in the asymptotically flat inertial region while Bob and Charlie hover near the event horizon, and their field modes undergo the Hawking--Unruh Bogoliubov transformation. 

The remainder of this paper is organized as follows. Section II reviews the Schwarzschild spacetime background and the Hawking--Unruh Bogoliubov transformation for bosonic field modes. In Sec. III, we introduce the definition of R\'enyi-2 entropy and tripartite mutual information for Gaussian states. Section IV derives the information balance relation between physically accessible and inaccessible sectors. Numerical results and physical discussions are presented in Sec. V. Finally, Sec. VI summarizes our conclusions.

\section{Vacuum Structure of a Scalar Field in Schwarzschild Spacetime}

The Schwarzschild black hole is the first static solution based on Einstein's equations of general relativity. In
order to better understand the nature of the curved spacetime,
it is required to figure out the properties of its quantumness.
Firstly, we briefly review the structure of scalar field in Schwarzschild spacetime and quantization of the vacuum state~\cite{Fulling1973,FHawking1975,Davies1975,Unruh1976,BirrellDavies1982,Takagi1986,CrispinoHiguchiMatsas2008,Wald1994}. The metric of a Schwarzschild black hole with mass $M$ can be described as, 
\begin{equation}\label{eq:SchwarzschildMetric}
\begin{split}
ds^{2} = &-\left(1-\frac{2M}{r}\right)dt^{2}
+\left(1-\frac{2M}{r}\right)^{-1}dr^{2}  \\
&+r^{2}\left(d\theta^{2}+\sin^{2}\theta d\phi^{2}\right).
\end{split}
\end{equation}
A massless scalar field $\Phi$ satisfies the Klein-Gordon equation
\begin{equation}
\frac{1}{\sqrt{-g}}\partial_{\mu}
\left(\sqrt{-g}g^{\mu\nu}\partial_{\nu}\Phi\right)=0.
\label{eq:KleinGordon}
\end{equation}
Employing the separation
\[
\Phi_{\omega lm}=r^{-1}\chi_{\omega l}(r)Y_{lm}(\theta,\phi)e^{-i\omega t},
\]
one obtains the usual radial equation. Introducing the tortoise coordinate
\[
r_{*}=r+2M\ln\left|\frac{r}{2M}-1\right|
\]
and the retarded time $\mu=t-r_{*}$, the outgoing modes outside and inside the event horizon $r_{+}=2M$ have the asymptotic forms
\begin{align}
  \Psi^{\mathrm{out}}_{\omega lm}(r>r_{+})
  &\approx \frac{1}{\sqrt{4\pi\omega}\,r}
  e^{-i\omega \mu}Y_{lm}(\theta,\phi), \\
  \Psi^{\mathrm{out}}_{\omega lm}(r<r_{+})
  &\approx \frac{1}{\sqrt{4\pi\omega}\,r}
  e^{i\omega \mu}Y_{lm}(\theta,\phi).
\end{align}

Using these exterior and interior Schwarzschild modes, the outgoing field can be expanded as

\begin{align}
\Phi_{\mathrm{out}} = & \smashoperator{\sum_{lm}}\int\!d\omega\kern-2pt
\Bigl[
b^{\mathrm{in}}_{\omega lm}\Psi^{\mathrm{out}}_{\omega lm}(r<r_{+})
+b^{\mathrm{in}\,\dagger}_{\omega lm}
\Psi^{\mathrm{out}\,*}_{\omega lm}(r<r_{+}) \nonumber \\
& + b^{\mathrm{out}}_{\omega lm}\Psi^{\mathrm{out}}_{\omega lm}(r>r_{+})
+b^{\mathrm{out}\,\dagger}_{\omega lm}
\Psi^{\mathrm{out}\,*}_{\omega lm}(r>r_{+}\kern-1pt)
\Bigr],
\label{eq:fieldExpansion}
\end{align}

where $b^{\mathrm{in}}_{\omega lm}$ and $b^{\mathrm{out}}_{\omega lm}$ annihilate the interior and exterior Schwarzschild modes, respectively.

The Kruskal vacuum is related to the Schwarzschild vacuum by a Bogoliubov transformation~\cite{FHawking1975,BirrellDavies1982,CrispinoHiguchiMatsas2008}. For each frequency mode, it can be written as a two-mode squeezed state of the exterior and interior Schwarzschild modes~\cite{BirrellDavies1982,AdessoFuentesEricsson2007,WangCaoJingFan2016},
\begin{equation}
\begin{split}
|0_{\omega}\rangle_{K}
&= \frac{1}{\cosh\eta}
\sum_{n=0}^{\infty}\tanh^{n}\eta\,
|n_{\omega}\rangle_{\mathrm{out}}
|n_{\omega}\rangle_{\mathrm{in}}  \\
&= \hat{U}_{\mathrm{out,in}}(\eta)
|0\rangle_{\mathrm{out}}|0\rangle_{\mathrm{in}} .
\end{split}
\label{eq:KruskalVacuum}
\end{equation}
Here
\[
\hat{U}_{\mathrm{out,in}}(\eta)
=
\exp\left[
\eta\left(
b_{\mathrm{out}}^{\dagger}b_{\mathrm{in}}^{\dagger}
-
b_{\mathrm{out}}b_{\mathrm{in}}
\right)
\right]
\]
is the two-mode squeezing operator. In the Hawking--Unruh correspondence, the same squeezing parameter can be expressed in terms of the Unruh temperature $T$ as
\begin{equation}
\cosh\eta=\frac{1}{\sqrt{1-e^{-\omega/T}}}.
\label{eq:eta_temperature}
\end{equation}
Equivalently, for a uniformly accelerated observer with acceleration $a$, one has $T=a/(2\pi)$.

In phase space, the two-mode squeezing operation is represented by the symplectic matrix~\cite{BraunsteinVanLoock2005,Weedbrook2012,AdessoFuentesEricsson2007,WangCaoJingFan2016}
\begin{equation}
S_{\mathrm{out,in}}(\eta) = \begin{pmatrix}
\cosh\eta & 0 & \sinh\eta & 0\\
0 & \cosh\eta & 0 & -\sinh\eta\\
\sinh\eta & 0 & \cosh\eta & 0\\
0 & -\sinh\eta & 0 & \cosh\eta
\end{pmatrix}.
\label{eq:SymplecticHawking}
\end{equation}

\section{QUANTUM MUTUAL INFORMATION IN SCHWARZSCHILD SPACETIME}

In this section, we focus on the dynamical evolution of quantum mutual information for Gaussian states in curved spacetime, revealing how Hawking radiation affects quantum correlations in continuous variable quantum systems.

\subsection{Quantum mutual information for Gaussian states}

We first go over some notations of the Gaussian state used in this work. An $n$-mode continuous-variable system is described by the tensor product Hilbert space $\mathcal{H}=\otimes_{i=1}^{n}\mathcal{H}_i$~\cite{BraunsteinVanLoock2005,Weedbrook2012,AdessoRagyLee2014,Serafini2017}. For each mode, one can  define the quadrature operators
\[
\hat{x}_i=a_i+a_i^{\dagger},\qquad
\hat{p}_i=-\mathrm{i}(a_i-a_i^{\dagger}),
\]
and collect them into a vector
\[
\hat{R}=(\hat{x}_1,\hat{p}_1,\cdots,\hat{x}_n,\hat{p}_n).
\]
The elements of vector $\hat{R}$ satisfy the canonical relation below
\begin{equation}
[\hat{R}_i,\hat{R}_j]=2\mathrm{i}\Omega_{ij}.
\end{equation}
Here, we can construct a standard symplectic form with entries  \(\Omega_{ij}\),  namely,
\begin{equation}
\boldsymbol{\Omega}
=(\Omega_{ij})=
\bigoplus_{i=1}^{n}
\begin{pmatrix}
0 & 1\\
-1 & 0
\end{pmatrix}.
\end{equation}

A Gaussian state $\rho$ is fully specified by its first moments and covariance matrix. Since first moments can be changed by local displacements and do not affect the correlation quantities considered here, we set them to zero. The covariance matrix $\sigma$ with elements $\sigma_{ij}$ is then defined by
\begin{equation}
\sigma_{ij}
=
\frac{1}{2}
\langle \hat{R}_i\hat{R}_j+\hat{R}_j\hat{R}_i\rangle
-
\langle \hat{R}_i\rangle\langle \hat{R}_j\rangle,
\end{equation}
which is a real symmetric matrix satisfying the uncertainty relation
\begin{equation}
\sigma+\mathrm{i}\boldsymbol{\Omega}\geq 0.
\end{equation}
 A Gaussian state is pure if and only if $\det\sigma=1$.

For continuous  quantum systems, the R\'enyi-2 entropy serves as a an important tool for characterizing information carried by the system, and it can be expressed in terms of the covariance matrix,
\begin{equation}
H_2(\rho)=-\ln\mathrm{Tr}(\rho^2)
=\frac{1}{2}\ln\det\sigma .
\end{equation}
We thus adopt it to evaluate mutual information for continuous quantum systems~\cite{Adesso2012}.
The bipartite mutual information for a two-mode Gaussian state $\rho_{AB}$ can be expressed as
\begin{equation}\label{defbipartite}
\begin{split}
I(A:B)
&=H_2(\rho_A)+H_2(\rho_B)-H_2(\rho_{AB})\\
&=\frac{1}{2}
\left(
\ln\det\sigma_A
+\ln\det\sigma_B
-\ln\det\sigma_{AB}
\right).
\end{split}
\end{equation}
where $\sigma_A$, $\sigma_B$ denote the covariance matrices of reduced states $\rho_A$, $\rho_B$, respectively. It is nonnegative, vanishes iff $\rho_{AB}=\rho_A\otimes\rho_B$, is invariant under local unitary operations, and satisfies the data-processing inequality under local quantum channels~\cite{NielsenChuang2010,Wilde2013,GroismanPopescuWinter2005}, making it a faithful measure of total (classical and quantum) bipartite correlations.

Since tripartite mutual information can be expressed in terms of bipartite mutual information~\cite{CerfAdami1997,Kumar2017},
\begin{equation}
I(A:B:C)=I(A:B)+I(A:C)-I(A:BC).
\label{eq:tmi_bipartite_decomposition}
\end{equation}
 One can extend it to three-mode Gaussian states $\rho_{ABC}$ in the inclusion-exclusion form~\cite{CerfAdami1997,Kumar2017}, that is,
\begin{equation}
\begin{aligned}
I(A:B:C)
         &= \frac{1}{2}\Big( \ln\det\sigma_A + \ln\det\sigma_B + \ln\det\sigma_C \\
         &\quad - \ln\det\sigma_{AB} - \ln\det\sigma_{AC}  - \ln\det\sigma_{BC}\\
         &\quad+ \ln\det\sigma_{ABC} \Big).
\end{aligned}
\label{eq:TMIdef}
\end{equation}

Eq.(\ref{eq:TMIdef}) gives a concise measure for the mutual information of three-mode continuous variable quantum systems, which is fully determined by the covariance matrices of the Gaussian state and its subsystems. If the mutual information \(I(A:B:C)=0\), the three-mode Gaussian state is separable; otherwise, it is entangled.

\subsection{Dynamical evolution of the quantum mutual information}

We consider a pure three-mode entangled Gaussian state $\rho_{ABC}$ shared by three inertial observers, Alice ($A$), Bob ($B$), and Charlie ($C$). The initial state is taken to be completely symmetric and is characterized by a global squeezing parameter $s\geq 0$~\cite{BraunsteinVanLoock2005,Weedbrook2012,AdessoIlluminati2007,AdessoSerafiniIlluminati2006}, and its covariance matrix takes the following block-matrix form,
\begin{equation}
\sigma_{ABC}^{(\mathrm{ini})}(s)=
\begin{pmatrix}
\sigma_A(s) & \zeta(s) & \zeta(s) \\
\zeta(s) & \sigma_B(s) & \zeta(s) \\
\zeta(s) & \zeta(s) & \sigma_C(s)
\end{pmatrix},
\label{eq:sigmaABC}
\end{equation}
Here,
\[
\sigma_A(s)=\sigma_B(s)=\sigma_C(s)=\mathrm{diag}(b,b),
\quad
\zeta(s)=\mathrm{diag}(z_1,z_2),
\]
and $b$, $z_1$, $z_2$  denote variables depending on the squeezing parameter $s$,
\begin{align}
b &=\frac{1}{3}\sqrt{4\cosh(4s)+5},\nonumber\\[1mm]
z_1 &=\frac{2\sinh^2(2s)+3\sinh(4s)}
{3\sqrt{4\cosh(4s)+5}},\nonumber\\[1mm]
z_2 &=\frac{2\sinh^2(2s)-3\sinh(4s)}
{3\sqrt{4\cosh(4s)+5}}.
\label{eq:parameters}
\end{align}
This Gaussian state can be generated, in principle, by sending single-mode squeezed vacua through a symmetric linear-optical beam-splitter network~\cite{VanLoockBraunstein2000}.

\begin{figure}[t]
\centering
\includegraphics[width=\columnwidth]{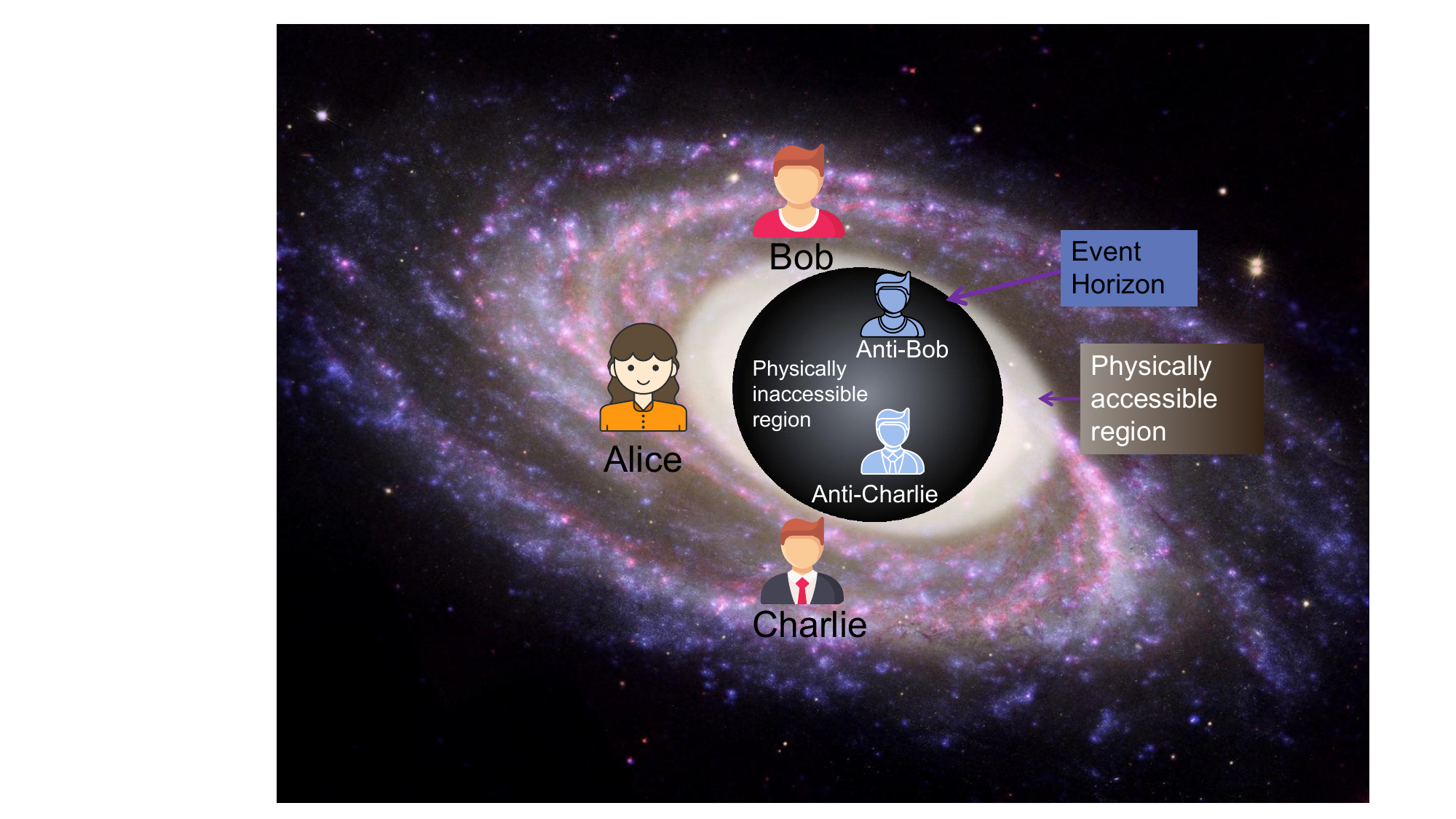}
\caption{Schematic illustration of the mode decomposition across the Schwarzschild event horizon. Alice's mode $A$ remains outside the horizon, whereas Bob's and Charlie's modes are transformed by the Hawking--Unruh Bogoliubov transformation into exterior modes $B_I$ and $C_I$ and interior modes $B_{II}$ and $C_{II}$. The exterior modes constitute the physically accessible sector, while the interior modes are physically inaccessible to an exterior observer.}
\label{fig:scenario_inaccessible}
\end{figure}

Now we assume Alice remains inertial, whereas Bob and Charlie hover near the event horizon. According to the Bogoliubov transformation, Bob's and Charlie's modes are split into exterior and interior components across the horizon as shown in Fig.\ref{fig:scenario_inaccessible}.  Specifically, they are mapped to physically accessible mode $B_I$ and $C_I$, physically inaccessible mode $B_{II}$ and $C_{II}$, respectively. Therefore, the initial three-mode Gaussian state is evolved into a five-mode Gaussian state
\(\rho_{A B_I B_{II} C_I C_{II}}\) with covariance matrix 
\begin{multline}\label{evostate}
\sigma_{AB_{\mathrm{I}}B_{\mathrm{II}}C_{\mathrm{I}}C_{\mathrm{II}}}(s) 
= \bigl[I_A \oplus S_{B_{\mathrm{I}}B_{\mathrm{II}}}(\eta_B) \oplus S_{C_{\mathrm{I}}C_{\mathrm{II}}}(\eta_C)\bigr] \times \\
\bigl[\sigma_{ABC}^{(\mathrm{ini})}(s) \oplus I_{B_{\mathrm{II}}} \oplus I_{C_{\mathrm{II}}}\bigr] \bigl[I_A \oplus S_{B_{\mathrm{I}}B_{\mathrm{II}}}(\eta_B) \oplus S_{C_{\mathrm{I}}C_{\mathrm{II}}}(\eta_C)\bigr]^{\mathrm{t}}.
\end{multline}

Substitute Eqs. (\ref{eq:SymplecticHawking}) and (\ref{eq:sigmaABC}) into the above formula to obtain the covariance matrix of the five-mode quantum state $\rho_{AB_IB_{II}C_IC_{II}}$ evolved via the Unruh effect.
Since region I outside the black hole and region II inside are disconnected, we should trace out the interior modes $B_{\text{II}}$ and $C_{\text{II}}$. Then, employing Eq.(\ref{eq:TMIdef}), the quantum mutual information of reduced state $\rho_{A B_{\text{I}} C_{\text{I}}}$ can be obtained,
\begingroup
\small
\begin{equation}
\begin{aligned}
&I(A:B_{\text{I}}:C_{\text{I}}) = \ln b+\ln d_{B_{\text{I}}}+\ln d_{C_{\text{I}}} + \\
&\frac{1}{2}\ln\Bigl[
\bigl(bd_{B_{\text{I}}}d_{C_{\text{I}}}+2z_1^3c_B^2c_C^2
-z_1^2(bc_B^2c_C^2+c_B^2d_{C_{\text{I}}}+c_C^2d_{B_{\text{I}}})\bigr)\\
&\times
\bigl(bd_{B_{\text{I}}}d_{C_{\text{I}}}+2z_2^3c_B^2c_C^2
-z_2^2(bc_B^2c_C^2+c_B^2d_{C_{\text{I}}}+c_C^2d_{B_{\text{I}}})\bigr)
\Bigr]\\
&-\frac{1}{2}\ln\Bigl[
(d_{B_{\text{I}}}d_{C_{\text{I}}}-z_1^2c_B^2c_C^2)
(d_{B_{\text{I}}}d_{C_{\text{I}}}-z_2^2c_B^2c_C^2)
\Bigr]\\
&-\frac{1}{2}\ln\Bigl[
(bd_{B_{\text{I}}}-z_1^2c_B^2)(bd_{B_{\text{I}}}-z_2^2c_B^2)
\Bigr]\\
&-\frac{1}{2}\ln\Bigl[
(bd_{C_{\text{I}}}-z_1^2c_C^2)(bd_{C_{\text{I}}}-z_2^2c_C^2)
\Bigr],
\end{aligned}
\label{eq:accessible}
\end{equation}
\endgroup
where $d_{i_I}=b\,c_i^2+s_i^2$, $d_{i_{II}}=b\,s_i^2+c_i^2$, 
$c_i=\cosh\eta_i$, 
$s_i=\sinh\eta_i$, 
$ i\in\{B,C\}$.

Exactly, the quantum  mutual information of $\rho_{A B_{\text{I}} C_{\text{I}}}$ depends on the squeezing parameter $s$, Hawking temperature $T$, and frequency $\omega$. This implies that the degree of correlation in the quantum system depends on these three variables.

By a similar approach, the quantum mutual information of other tripartite reduced states  $\rho_{A B_{\text{II}} C_{\text{II}}}$,  $\rho_{A B_{\text{I}} C_{\text{II}}}$ and  $\rho_{A B_{\text{II}} C_{\text{I}}}$ also can be calculated. Furthermore, combining the definitions of mutual information for bipartite  Gaussian states given in 
Eq.~(\ref{defbipartite}), one can obtain the quantum mutual information of bipartite reduced states $\rho_{AB_I}$, $\rho_{AC_I}$, $\rho_{AB_{II}}$ and $\rho_{AC_{II}}$. Their expressions are rather involved and thus omitted here; for detailed derivations, see the Appendix \ref{app:analytical_expressions}.

\section{Tripartite Information Balance Across Causal Horizons}


Quantum correlations are generally not arbitrarily distributed among subsystems. Due to the nonlocality of quantum systems, their distribution is subject to certain constraints, exemplified by trade-off relations~\cite{CoffmanKunduWootters2000,OsborneVerstraete2006,HaydenHeadrickMaloney2013}. In black-hole settings, the redistribution of correlations between accessible and inaccessible sectors has been reported for entanglement, mutual information, and multipartite quantum resources under the Hawking effect~\cite{PanJing2008,WangPanJing2010,Wu2024}. 

Next, we  derive two useful relations for the  mutual information. 
Since the initial three-mode Gaussian state is pure,  i.e., $\det \sigma_{ABC}^{(\mathrm{ini})}(s) =1$,  and two-mode squeezing operater $S_{\mathrm{out,in}}(\eta)$ is unitary, it follows from Eq. (\ref{evostate}) that 
$\det\sigma_{AB_{\mathrm{I}}B_{\mathrm{II}}C_{\mathrm{I}}C_{\mathrm{II}}}(s)=1$.
That means the five-mode state $\rho_{A B_I B_{II} C_I C_{II}}$ is pure.

Additionally, for any  pure state $\rho_{X\Bar{X}}$,  the R\'enyi-2 entropies of  $\rho_X$ and $\rho_{\bar X}$ are equal ~\cite{NielsenChuang2010,Wilde2013} (a proof is given in Appendix~\ref{app:complementary_entropy}), namely,
\begin{equation}
H_2(\rho_X)=H_2(\rho_{\bar X}).
\label{eq:complementary_entropy_general}
\end{equation}

Therefore,  
the equalities as belows hold for five-mode pure state $\rho_{A B_I B_{II} C_I C_{II}}$,
\begin{align}
H_2(\rho_{A B_I C_I}) &= H_2(\rho_{B_{II} C_{II}}), \nonumber\\
H_2(\rho_{A B_{II} C_{II}}) &= H_2(\rho_{B_I C_I}), \nonumber\\
H_2(\rho_{A B_I C_{II}}) &= H_2(\rho_{B_{II} C_I}), \nonumber\\
H_2(\rho_{A B_{II} C_I}) &= H_2(\rho_{B_I C_{II}}). \label{eq:comp4}
\end{align}

According to the definition of bipartite mutual information given in Eq.~(\ref{defbipartite}), it is found that 
\begin{align}
&I(A:B_I C_I) + I(A:B_{II} C_{II}) \nonumber\\
&= 2H_2(\rho_A) + \big[H_2(\rho_{B_I C_I})+H_2(\rho_{B_{II} C_{II}})\big] \nonumber\\
&\quad - \big[H_2(\rho_{A B_I C_I})+H_2(\rho_{A B_{II} C_{II}})\big].
\end{align}
\begin{align}
&I(A:B_I C_{II}) + I(A:B_{II} C_I) \nonumber\\
&= 2H_2(\rho_A) + \big[H_2(\rho_{B_I C_{II}})+H_2(\rho_{B_{II} C_I})\big] \nonumber\\
&\quad - \big[H_2(\rho_{A B_I C_{II}})+H_2(\rho_{A B_{II} C_I})\big].
\end{align}
Substituting Eqs. (\ref{eq:comp4}) to the equations above, it can yield the relations as below,
\begin{align}
I(A:B_I C_I) + I(A:B_{II} C_{II}) &= 2H_2(\rho_A),  \\
I(A:B_I C_{II}) + I(A:B_{II} C_I) &= 2H_2(\rho_A). \label{eq:sumrule2}
\end{align}
Therefore,  it means the following equation hold 
\begin{equation}
\begin{split}
& I(A:B_I C_I) + I(A:B_{II} C_{II}) \\
={}& I(A:B_I C_{II}) + I(A:B_{II} C_I).
\end{split}
\label{eq:sumrule_const}
\end{equation}

\begin{figure}[t]
\centering
\includegraphics[width=\columnwidth]{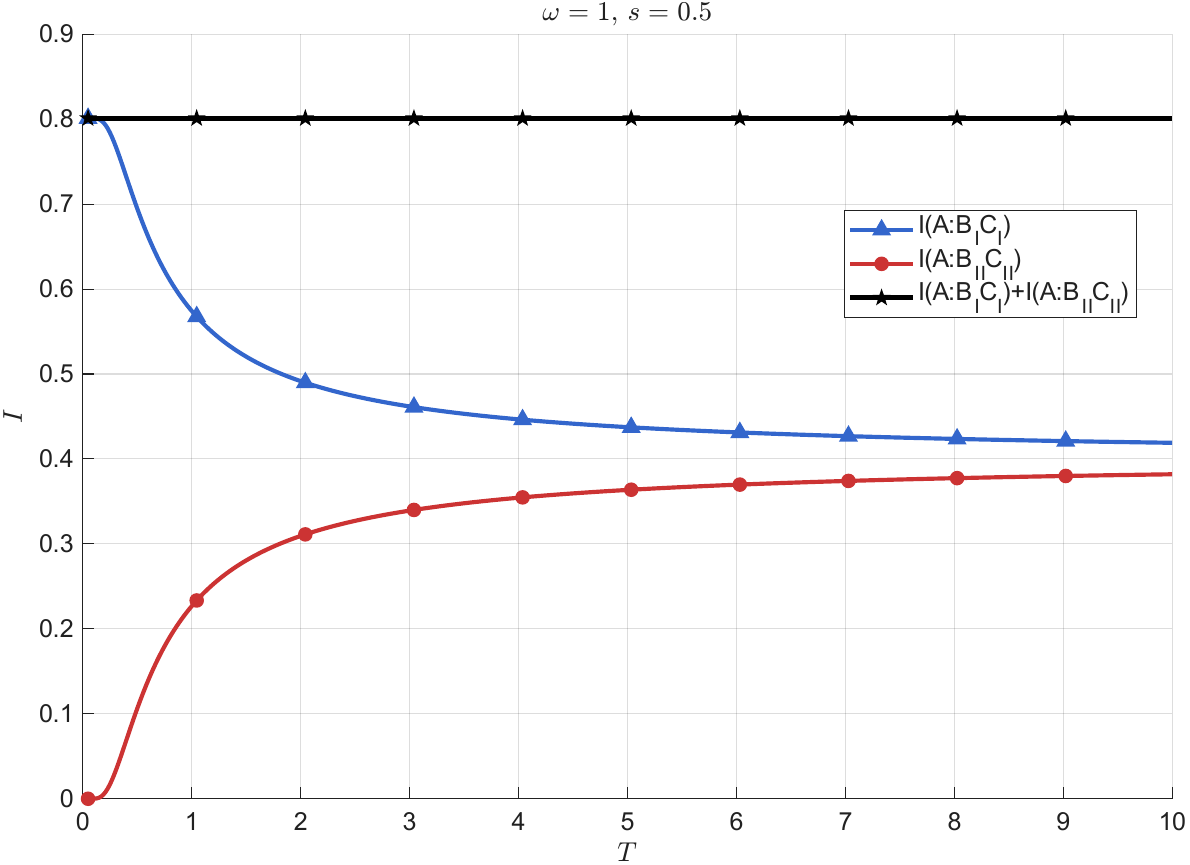}
\caption{Temperature dependence of the composite bipartite mutual informations between Alice and the two-mode exterior or interior sectors at fixed $\omega=1$ and $s=0.5$. The plotted quantities are $I(A:B_I C_I)$, $I(A:B_{II}C_{II})$, and their sum. The exterior contribution decreases with the Hawking temperature $T$, whereas the interior contribution increases. Their sum remains constant at $2\ln b\approx 0.80096$, in agreement with the sum rule in Eq.~(\ref{eq:sumrule_const}).
}
\label{fig:composite_bipartite_balance}
\end{figure}

Since the reduced state $\rho_A$ has covariance matrix $\sigma_A = diag(b, b)$, it yields 
$I(A:B_I C_I) + I(A:B_{II} C_{II})= 2\ln b$.
As shown in Fig.~\ref{fig:composite_bipartite_balance}, we plot function $I(A:B_I C_I)$ and $I(A:B_{II} C_{II})$ for the parameters $\omega=1$ and $s=0.5$. In this case, we have $b = \frac{1}{3}\sqrt{4\cosh 2 + 5} \approx 1.492529$. As the Hawking temperature increases, the bipartite mutual information $I(A:B_I C_I)$ decreases while  $I(A:B_{II} C_{II})$ increases, yet their sum is invariant.  This behaviour clearly demonstrates that the mutual information is redistributed between the exterior and interior regions.

 To investigate how the tripartite mutual information are distributed among the different subsystems, we define two aggregate quantities:
\begin{equation}
\mathcal{I}_{\mathrm{inclusive}} \equiv I(A:B_I:C_I) + I(A:B_{II}:C_{II}), \label{eq:Iinclusive}
\end{equation}
and
\begin{equation}
\mathcal{I}_{\mathrm{mixed}} \equiv I(A:B_I:C_{II}) + I(A:B_{II}:C_I). \label{eq:Imixed}
\end{equation}

Using  the definition of tripartite mutual information as shown in Eq.~(\ref{eq:tmi_bipartite_decomposition}), it can be obtained 
\begin{align}
\mathcal{I}_{\mathrm{inclusive}} - \mathcal{I}_{\mathrm{mixed}}
&= \big[I(A:B_I C_{II}) + I(A:B_{II} C_I)\big] \nonumber\\
&\quad - \big[I(A:B_I C_I) + I(A:B_{II} C_{II})\big],
\end{align}
Substituting Eq.~(\ref{eq:sumrule_const}) into the above expressions, we immediately obtain
\begin{equation}
\begin{split}
& I(A:B_I:C_I) + I(A:B_{II}:C_{II}) \\
={}& I(A:B_I:C_{II}) + I(A:B_{II}:C_I).
\end{split}
\label{eq:tripartite_balance_explicit}
\end{equation}

Exactly, Eq.~\eqref{eq:tripartite_balance_explicit} reveals the distribution of mutual information for Gaussian states after the effect of the Unruh effect. Furthermore, it manifests the complementarity relation about mutual information between subsystems.
It is worth noting that Eqs.  (\ref{eq:sumrule_const} ) and (\ref{eq:tripartite_balance_explicit})  always hold for arbitrary three-mode Gaussian states, independent of the choice of each parameter of the initial state.

\section{Results and Discussion}
In this section, we systematically analyze the influence of the Unruh effect on quantum mutual information. Before presenting the numerical results, we emphasize that the behavior of the R\'enyi-2 tripartite mutual information should be distinguished from that of other quantum resources in related black-hole settings. Previous studies have shown that entanglement, Bell nonlocality, coherence, and Gaussian steering may exhibit degradation, sudden death, sudden birth, or redistribution between accessible and inaccessible sectors under Hawking radiation~\cite{PanJing2008,WangPanJing2010,XuSongShiYe2014PRD,XuSongShiYe2014PLB,DaiShenShi2016PRD,WangTianJingFan2016,WangCaoJingFan2016,AliAlKuwariGhominejad2024,Wu2025,LiuWenWang2026PLB}. Multipartite coherence and its distribution under the Unruh effect have also been investigated in related noninertial settings~\cite{WuLiZeng2021QIP,WuZengCao2021CQG}. In contrast, the quantity studied here characterizes a mutual-information-based correlation structure, and therefore its temperature dependence need not follow the same monotonic degradation pattern.

\subsection{Effect of Initial Squeezing $s$}

\begin{figure*}[t]
  \centering
  \begin{minipage}{0.32\textwidth}
    \centering
    \includegraphics[width=\linewidth]{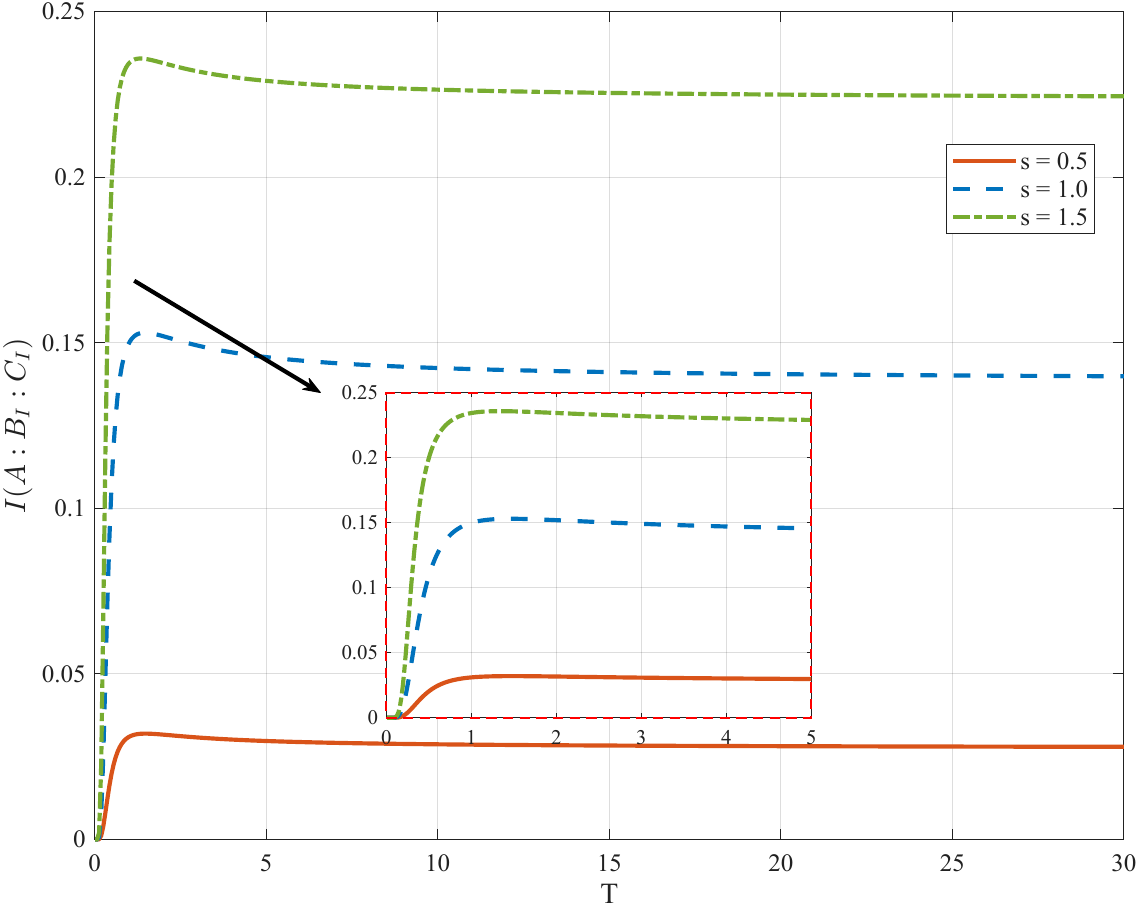}
    \centerline{(a) }
  \end{minipage}
  \hfill
  \begin{minipage}{0.32\textwidth}
    \centering
    \includegraphics[width=\linewidth]{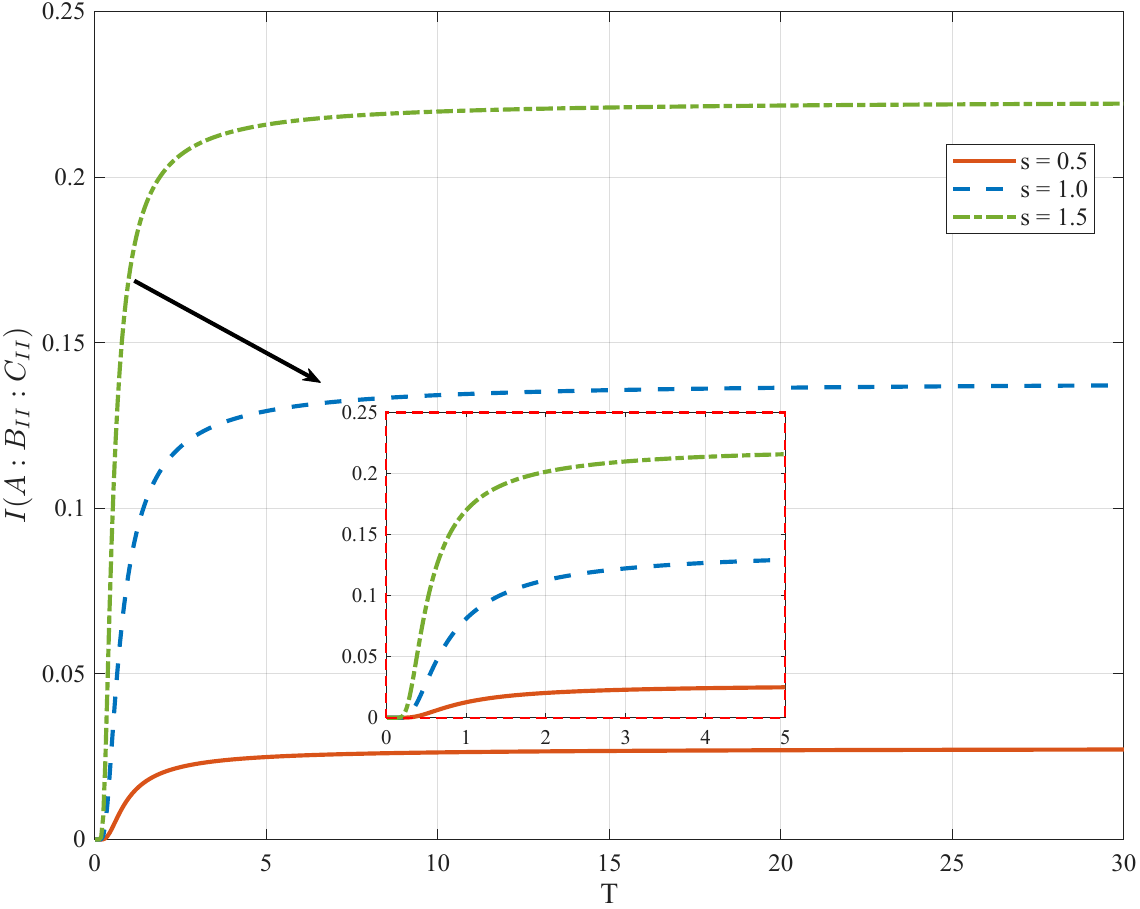}
    \centerline{(b) }
  \end{minipage}
  \hfill
  \begin{minipage}{0.32\textwidth}
    \centering
    \includegraphics[width=\linewidth]{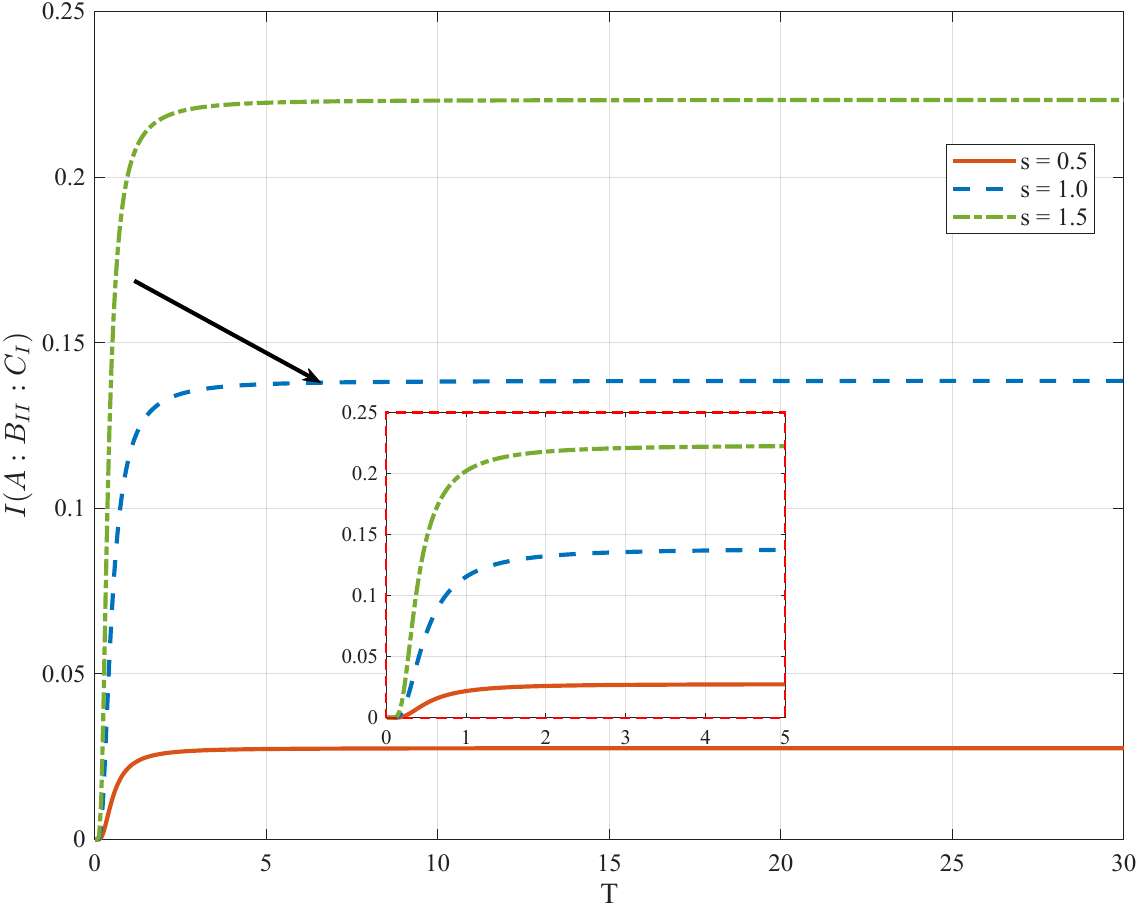}
    \centerline{(c) }
  \end{minipage}
  \caption{Tripartite mutual information as a function of the Hawking temperature $T$ for various initial squeezing strengths $s$ at a fixed mode frequency $\omega=1.0$. The panels display the corresponding values for three different partitions: (a) $I(A:B_I:C_I)$, (b) $I(A:B_{II}:C_{II})$, (c) $I(A:B_{II}:C_I)$. }  \label{fig:dynamics_squeezing}
\end{figure*}

Without loss of generality, 
 we consider the symmetric acceleration condition, i.e., $\eta_B=\eta_C=\eta$. 
As shown in  Fig.~\ref{fig:dynamics_squeezing}, two prominent phenomena can be observed. (i)The behavior of physically accessible mutual information $I(A:B_I:C_I)$  is non-monotonic: as the Hawking temperature increases, it first rises, then falls, and ultimately settles to a constant value; Moreover, no sudden death takes place, which differs from the sudden-death behavior reported for Gaussian steering under the Hawking effect in Schwarzschild spacetime~\cite{WangCaoJingFan2016,Wu2025}.
(ii)Physically inaccessible mutual information $I(A:B_{II}:C_{II})$ and $I(A:B_{II}:C_I)$ increase with temperature and eventually saturate.

Interestingly, the sharp change in both physically accessible and inaccessible mutual information takes place in the low temperature region, i.e., $T\in [0, 5]$,  while for temperatures above 5, the mutual information varies slowly and tends to stabilize.
Moreover, a higher squeezing strength $s$ is associated with a larger mutual information. It implies that a larger squeezing strength provides stronger quantum correlation for the  Gaussian states.

Furthermore, the non-monotonic behaviour of  $I(A:B_I:C_I)$ can be characterized by the critical temperature $T_{\mathrm{critical}}$, defined by
\begin{equation}
\left.
\frac{\partial I(A:B_I:C_I)}{\partial T}
\right|_{T=T_{\mathrm{critical}}}=0 .
\end{equation}

Table. \ref{tab:varying_s} shows the critical temperature at which quantum mutual information $I(A:B_I:C_I)$ reaches its maximum varies with the squeezing strength. The greater the squeezing strength, the lower the critical temperature.
\begin{table}[t]
\begin{ruledtabular}
\begin{tabular}{lcc}
 $s$ & $I_{\text{max}}$ &  $T_{\text{critical}}$ \\
\colrule
0.5 & 0.0320 & 1.48 \\
1.0 & 0.1530 & 1.46 \\
1.5 & 0.2358 & 1.33 \\
\end{tabular}
\end{ruledtabular}
\caption{Peak values of the accessible tripartite information $I(A:B_I:C_I)$ for different initial squeezing strengths $s$ at fixed $\omega=1.0$. The peak value increases with $s$, whereas the critical temperature shows a small downward shift.}
\label{tab:varying_s}
\end{table}

\subsection{Effect of Mode Frequency $\omega$}

This subsection is devoted to the analysis of the mode-frequency dependence of the tripartite mutual information, with the initial squeezing strength fixed at $s=0.5$ under the symmetric acceleration condition $\eta_B=\eta_C=\eta$.

\begin{figure*}[t]
  \centering
  \begin{minipage}{0.32\textwidth}
    \centering
    \includegraphics[width=\linewidth]{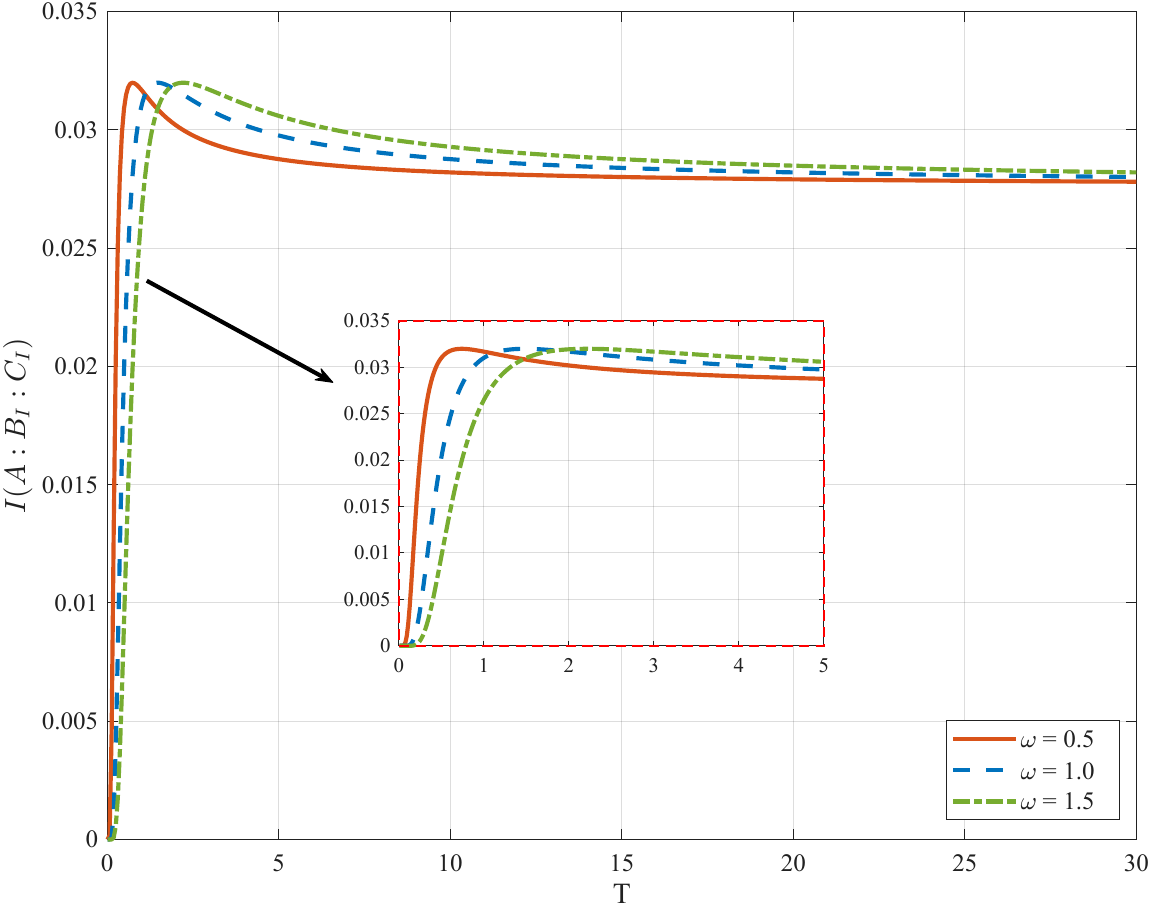}
    \centerline{(a)}
  \end{minipage}
  \hfill
  \begin{minipage}{0.32\textwidth}
    \centering
    \includegraphics[width=\linewidth]{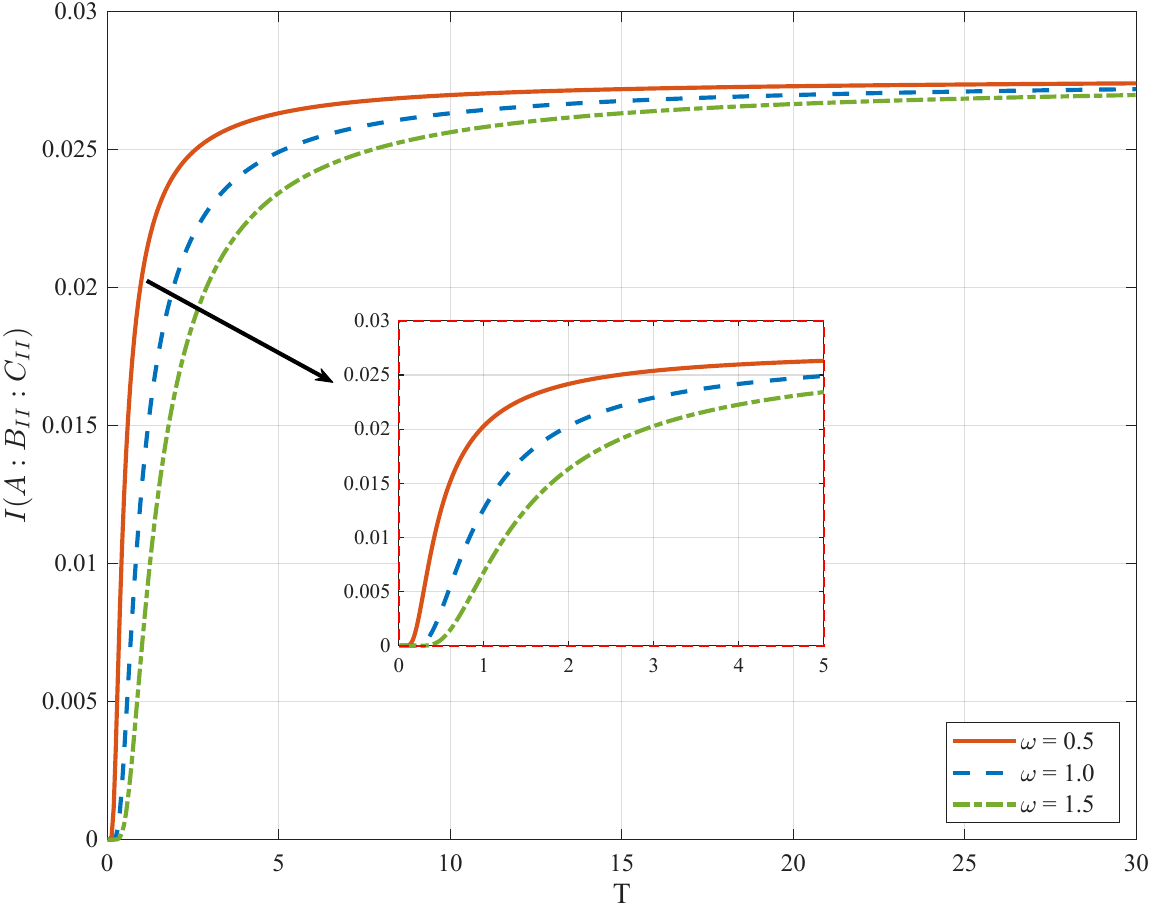}
    \centerline{(b)}
  \end{minipage}
  \hfill
  \begin{minipage}{0.32\textwidth}
    \centering
    \includegraphics[width=\linewidth]{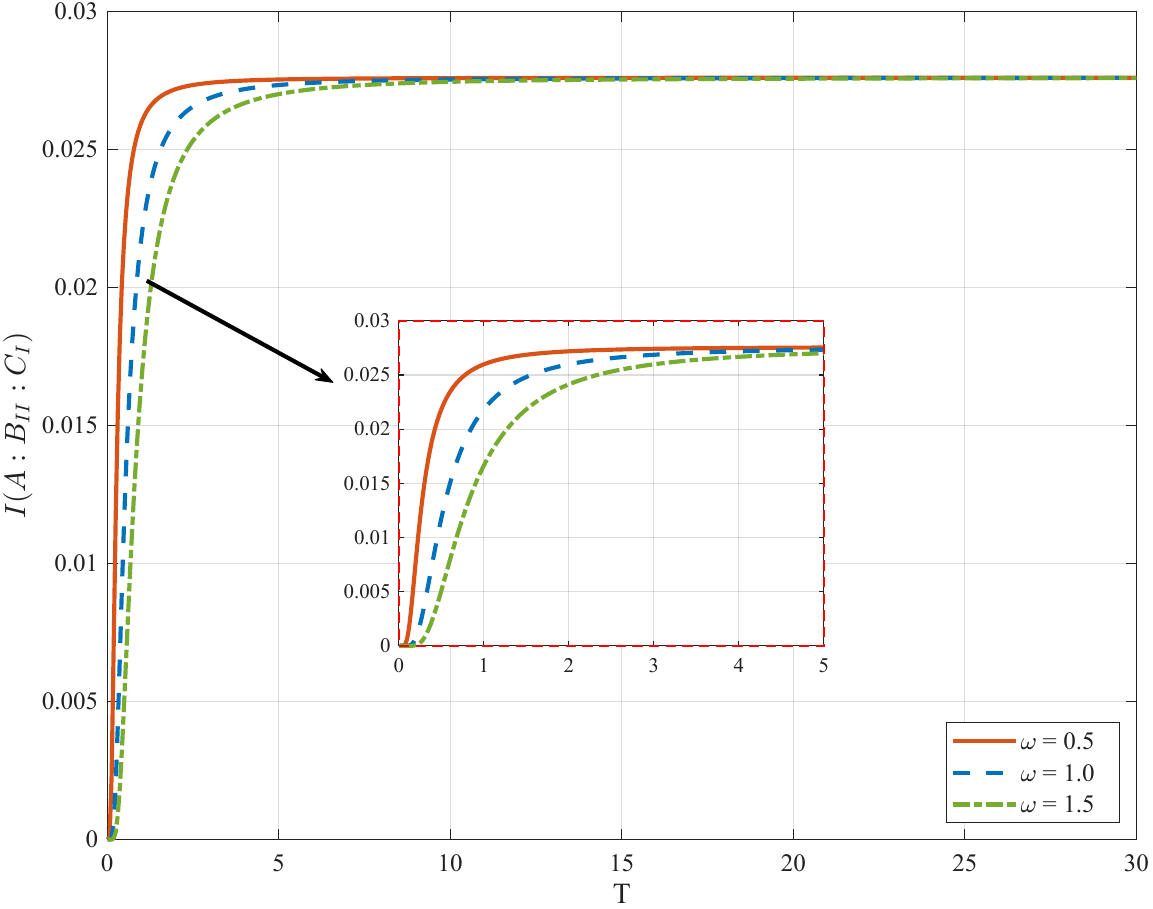}
    \centerline{(c)}
  \end{minipage}
  \caption{Tripartite mutual information as a function of the Hawking temperature $T$ for different mode frequencies $\omega$ at a fixed initial squeezing strength $s=0.5$. The panels display the corresponding values for three different partitions: (a) $I(A:B_I:C_I)$, (b) $I(A:B_{II}:C_{II})$, and (c) $I(A:B_{II}:C_I)$.}
  \label{fig:dynamics_frequency}
\end{figure*}

The mode frequency $\omega$ has a significant influence on  the tripartite mutual information. As shown in Fig.~\ref{fig:dynamics_frequency}, the physically accessible mutual information $I(A:B_I:C_I)$ exhibits a non-monotonic behaviour: with increasing Hawking temperature, it first increases, reaches a maximum value, and then decreases toward a stable value.
Additionally, it is noteworthy that the maximum value of $I(A:B_I:C_I)$  is independent of $\omega$ and remains identical.
However, 
prior to attaining its maximum, $I(A:B_I:C_I)$ increases with decreasing $\omega$; beyond the maximum, the trend reverses.
For the physically inaccessible mutual information $I(A:B_{II}:C_{II})$ and $I(A:B_{II}:C_I)$, they increase monotonically with the Hawking temperature and finally approach saturation.

Similarly, the non-monotonic behaviour of $I(A:B_I:C_I)$ can also be characterized by the critical temperature $T_{\mathrm{critical}}$.
As presented in Table~\ref{tab:varying_omega}, the critical temperature at which $I(A:B_I:C_I)$ attains its maximum is computed for various values of $\omega$.
 It implies the mode frequency does not affects the maximal value of the accessible mutual information, but influences the corresponding critical temperature.

\begin{table}[t]
\begin{ruledtabular}
\begin{tabular}{lcc}
 $\omega$ & $I_{\text{max}}$ & $T_{\text{critical}}$ \\
\colrule
0.5 & 0.0320 & 0.74 \\
1.0 & 0.0320 & 1.48 \\
1.5 & 0.0320 & 2.21 \\
\end{tabular}
\end{ruledtabular}
\caption{Peak values of the accessible tripartite information $I(A:B_I:C_I)$ for different mode frequencies $\omega$ at fixed $s=0.5$. The peak value remains almost unchanged, whereas the critical temperature increases with $\omega$.}
\label{tab:varying_omega}
\end{table}

\section{Conclusion}

We have investigated the dynamical evolution of the quantum mutual information for a three-mode squeezed Gaussian state in Schwarzschild spacetime. Our analysis reveals that the physically inaccessible mutual information increases with the Hawking temperature and eventually saturates. In contrast, the physically accessible mutual information exhibits a nonmonotonic temperature dependence: it first rises, reaches a maximum value at a critical temperature, and subsequently declines toward a finite asymptotic value. This behavior is in marked contrast to the typical monotonic degradation patterns of entanglement and quantum coherence in Schwarzschild spacetime.

Furthermore, we systematically analyzed how the squeezing parameter $s$ and mode frequency $\omega$ affect the mutual information. A larger squeezing parameter $s$ leads to a larger mutual information and a lower critical temperature for the physically accessible partition. Conversely, the mode frequency $\omega$ does not affect the maximal value of the physically accessible mutual information, but a larger $\omega$ results in a higher critical temperature.

In addition, we established several constraint relations that govern the redistribution of mutual information among different subsystems. Since these relations strictly depend on the purity of the global state and the unitarity of the Bogoliubov transformation, they are universally independent of both the squeezing parameter and the mode frequency.

In summary, the Hawking--Unruh transformation in Schwarzschild spacetime serves as a mechanism for the redistribution, rather than the monotonic degradation, of continuous variable multipartite correlations. Extending the present analysis to more general curved spacetime backgrounds provides a natural framework for future investigations.

\section*{ACKNOWLEDGMENTS}
This work is supported by the Natural Science Foundation of Hainan Province under Grant No. 125RC744; the China Scholarship Council (CSC).

\appendix
\section{Proof of the complementary R\'enyi-2 entropy identities}
\label{app:complementary_entropy}

In this appendix, we justify the complementary entropy identities used in Eq.~(\ref{eq:comp4}). The initial three-mode Gaussian state $\rho_{ABC}$ considered in the main text is pure. The ancillary modes $B_{II}$ and $C_{II}$ are introduced as vacuum modes and are therefore pure. Hence the state before the Hawking--Unruh transformation,
\begin{equation}
\rho_{ABC}\otimes |0\rangle_{B_{II}}\langle 0|
\otimes |0\rangle_{C_{II}}\langle 0|,
\end{equation}
is pure. The Hawking--Unruh Bogoliubov transformation is implemented by two-mode squeezing unitary operations, or equivalently by symplectic transformations at the covariance-matrix level. Therefore, the purity of the global state is preserved, and the resulting five-mode state
$\rho_{A B_I B_{II} C_I C_{II}}$
is a pure state.

We now recall a standard consequence of the Schmidt decomposition. For a pure bipartite state $|\Psi\rangle_{X\bar X}$, where $\bar X$ denotes the complement of $X$, one can write
\begin{equation}
|\Psi\rangle_{X\bar X}
=
\sum_k \sqrt{\lambda_k}\,
|k\rangle_X |k\rangle_{\bar X},
\end{equation}
with $\lambda_k\geq 0$ and $\sum_k\lambda_k=1$. Taking the partial trace over either subsystem gives
\begin{equation}
\rho_X
=
\mathrm{Tr}_{\bar X}(|\Psi\rangle\langle\Psi|)
=
\sum_k \lambda_k |k\rangle_X\langle k|,
\end{equation}
and
\begin{equation}
\rho_{\bar X}
=
\mathrm{Tr}_{X}(|\Psi\rangle\langle\Psi|)
=
\sum_k \lambda_k |k\rangle_{\bar X}\langle k|.
\end{equation}
Thus, $\rho_X$ and $\rho_{\bar X}$ have the same nonzero eigenvalues. Although these reduced states are generally mixed, their R\'enyi-2 entropies are equal:
\begin{equation}
\mathrm{Tr}(\rho_X^2)
=
\sum_k \lambda_k^2
=
\mathrm{Tr}(\rho_{\bar X}^2),
\end{equation}
and hence
\begin{equation}
H_2(\rho_X)
=
-\ln\mathrm{Tr}(\rho_X^2)
=
-\ln\mathrm{Tr}(\rho_{\bar X}^2)
=
H_2(\rho_{\bar X}).
\end{equation}
In the Gaussian-state formulation, this is equivalently reflected by the fact that complementary subsystems of a pure Gaussian state have the same nontrivial symplectic spectrum; possible additional trivial symplectic eigenvalues are equal to unity and do not change the R\'enyi-2 entropy.

Applying this property to the four complementary bipartitions of the five-mode pure state,
\begin{align}
A B_I C_I \,|\, B_{II} C_{II}, \qquad
A B_{II} C_{II} \,|\, B_I C_I, \nonumber\\
A B_I C_{II} \,|\, B_{II} C_I, \qquad
A B_{II} C_I \,|\, B_I C_{II},
\end{align}
directly yields Eq.~(\ref{eq:comp4}).

\section{Analytical expressions for the remaining causal partitions}
\label{app:analytical_expressions}

The tripartite mutual information for the physically inaccessible partition is
\begin{align}
& I(A:B_{II}:C_{II}) = \ln b+\ln d_{B_{II}}+\ln d_{C_{II}} \nonumber \\
& + \frac{1}{2}\ln\Big[
\bigl(bd_{B_{II}}d_{C_{II}}+2z_1^3s_B^2s_C^2 \nonumber \\
& \quad - z_1^2(bs_B^2s_C^2+s_B^2d_{C_{II}}+s_C^2d_{B_{II}})\bigr) \times \nonumber \\
& \bigl(bd_{B_{II}}d_{C_{II}}+2z_2^3s_B^2s_C^2 \nonumber \\
& \quad - z_2^2(bs_B^2s_C^2+s_B^2d_{C_{II}}+s_C^2d_{B_{II}})\bigr)
\Big] \nonumber \\
& - \frac{1}{2}\ln\Big[
(d_{B_{II}}d_{C_{II}}-z_1^2s_B^2s_C^2)
(d_{B_{II}}d_{C_{II}}-z_2^2s_B^2s_C^2)
\Big] \nonumber \\
& - \frac{1}{2}\ln\Big[
(bd_{B_{II}}-z_1^2s_B^2)(bd_{B_{II}}-z_2^2s_B^2)
\Big] \nonumber \\
& - \frac{1}{2}\ln\Big[
(bd_{C_{II}}-z_1^2s_C^2)(bd_{C_{II}}-z_2^2s_C^2)
\Big] \kern-35pt .
\end{align}

\begin{align}
& I(A:B_{II}:C_I) = \ln b+\ln d_{B_{II}}+\ln d_{C_I} \nonumber \\
& + \frac{1}{2}\ln\Big[
\bigl(bd_{B_{II}}d_{C_I}+2z_1^3s_B^2c_C^2 \nonumber \\
& \quad - z_1^2(bs_B^2c_C^2+s_B^2d_{C_I}+c_C^2d_{B_{II}})\bigr) \times \nonumber \\
& \bigl(bd_{B_{II}}d_{C_I}+2z_2^3s_B^2c_C^2 \nonumber \\
& \quad - z_2^2(bs_B^2c_C^2+s_B^2d_{C_I}+c_C^2d_{B_{II}})\bigr)
\Big] \nonumber \\
& - \frac{1}{2}\ln\Big[
(d_{B_{II}}d_{C_I}-z_1^2s_B^2c_C^2)
(d_{B_{II}}d_{C_I}-z_2^2s_B^2c_C^2)
\Big] \nonumber \\
& - \frac{1}{2}\ln\Big[
(bd_{B_{II}}-z_1^2s_B^2)(bd_{B_{II}}-z_2^2s_B^2)
\Big] \nonumber \\
& - \frac{1}{2}\ln\Big[
(bd_{C_I}-z_1^2c_C^2)(bd_{C_I}-z_2^2c_C^2)
\Big] \kern-18pt .
\end{align}

\begin{align}
& I(A:B_I:C_{II}) = \ln b+\ln d_{B_I}+\ln d_{C_{II}} \nonumber \\
& + \frac{1}{2}\ln\Big[
\bigl(bd_{B_I}d_{C_{II}}+2z_1^3c_B^2s_C^2 \nonumber \\
& \quad - z_1^2(bc_B^2s_C^2+c_B^2d_{C_{II}}+s_C^2d_{B_I})\bigr) \times \nonumber \\
& \bigl(bd_{B_I}d_{C_{II}}+2z_2^3c_B^2s_C^2 \nonumber \\
& \quad - z_2^2(bc_B^2s_C^2+c_B^2d_{C_{II}}+s_C^2d_{B_I})\bigr)
\Big] \nonumber \\
& - \frac{1}{2}\ln\Big[
(d_{B_I}d_{C_{II}}-z_1^2c_B^2s_C^2)
(d_{B_I}d_{C_{II}}-z_2^2c_B^2s_C^2)
\Big] \nonumber \\
& - \frac{1}{2}\ln\Big[
(bd_{B_I}-z_1^2c_B^2)(bd_{B_I}-z_2^2c_B^2)
\Big] \nonumber \\
& - \frac{1}{2}\ln\Big[
(bd_{C_{II}}-z_1^2s_C^2)(bd_{C_{II}}-z_2^2s_C^2)
\Big] \kern-10pt .
\end{align}

The bipartite mutual information for the physically accessible partitions are
\begin{align}
I(A:B_I) &= \ln b + \ln d_{B_I} \nonumber \\
&\quad - \frac{1}{2}\ln \big[(b d_{B_I} - z_1^2 c_B^2)(b d_{B_I} - z_2^2 c_B^2)\big], \label{eq:I_ABI} \\[1.5ex]
I(A:C_I) &= \ln b + \ln d_{C_I} \nonumber \\
&\quad - \frac{1}{2}\ln \big[(b d_{C_I} - z_1^2 c_C^2)(b d_{C_I} - z_2^2 c_C^2)\big], \label{eq:I_ACI} \\[1.5ex]
I(A:B_I C_I) &= \ln b + \frac{1}{2}\ln \big[ (d_{B_I}d_{C_I} - z_1^2 c_B^2 c_C^2) \nonumber \\
&\qquad \times (d_{B_I}d_{C_I} - z_2^2 c_B^2 c_C^2) \big] \nonumber \\
&\quad - \frac{1}{2}\ln \Big[ \bigl(b d_{B_I}d_{C_I} - z_1^2 (b c_B^2 c_C^2 + c_B^2 d_{C_I} \nonumber \\
&\qquad + c_C^2 d_{B_I}) + 2z_1^3 c_B^2 c_C^2\bigr) \nonumber \\
&\qquad \times \bigl(b d_{B_I}d_{C_I} - z_2^2 (b c_B^2 c_C^2 + c_B^2 d_{C_I} \nonumber \\
&\qquad + c_C^2 d_{B_I}) + 2z_2^3 c_B^2 c_C^2\bigr) \Big].
\label{eq:I_ABICI}
\end{align}

The bipartite mutual information for the physically inaccessible partitions are
\begin{align}
& I(A:B_{II}) = \ln b + \ln d_{B_{II}} \nonumber \\
& \quad - \frac{1}{2}\ln \big[(b d_{B_{II}} - z_1^2 s_B^2)(b d_{B_{II}} - z_2^2 s_B^2)\big] \kern-5pt , \label{eq:I_ABII} \\
& I(A:C_{II}) = \ln b + \ln d_{C_{II}} \nonumber \\
& \quad - \frac{1}{2}\ln \big[(b d_{C_{II}} - z_1^2 s_C^2)(b d_{C_{II}} - z_2^2 s_C^2)\big] \kern-5pt , \label{eq:I_ACII} \\
& I(A:B_{II}C_{II}) = \ln b \nonumber \\
& + \frac{1}{2}\ln \big[ (d_{B_{II}}d_{C_{II}} - z_1^2 s_B^2 s_C^2)
(d_{B_{II}}d_{C_{II}} - z_2^2 s_B^2 s_C^2) \big] \nonumber \\
& - \frac{1}{2}\ln \Big[
 \bigl(b d_{B_{II}}d_{C_{II}} - z_1^2 (b s_B^2 s_C^2 + s_B^2 d_{C_{II}} + s_C^2 d_{B_{II}}) \nonumber \\
& \qquad + 2z_1^3 s_B^2 s_C^2\bigr) \times \nonumber \\
& \bigl(b d_{B_{II}}d_{C_{II}} - z_2^2 (b s_B^2 s_C^2 + s_B^2 d_{C_{II}} + s_C^2 d_{B_{II}}) \nonumber \\
& \qquad + 2z_2^3 s_B^2 s_C^2\bigr)
\Big] \kern-55pt \label{eq:I_ABIICII}
\end{align}
\vfill

\end{document}